\documentclass{PoS}

\title{Extracting strong phase and $CP$ violation in $D$ decays by using quantum correlations
in $\psi(3770)\to D^0 \overline{D}^0 \to (V_1V_2)(K \pi)$ and
$\psi(3770)\to D^0\overline{D}^0\to (V_1V_2)(V_3V_4)$}

\ShortTitle{Extracting $CP$ violation and strong phase in $D$ decays
by using quantum correlated double $D$ decay}

\author{\speaker{Xian-Wei Kang}\\
        Institute of High Energy Physics, P.O.Box 918, Beijing
100049, China\\ and\\ Department of Physics, Henan Normal
University, Xinxiang 453007, China\\
        E-mail: \email{kangxw@ihep.ac.cn}}
\author{J\'er\^ome Charles\\
       Centre de Physique Th\'eorique~\footnote{Laboratoire affili\'e \`a la FRUMAM},
CNRS \& Univ. Aix-Marseille 1 \& 2 and Sud Toulon-Var (UMR
6207), Luminy Case 907, 13288 Marseille Cedex 9, France\\
       E-mail: \email{charles@cpt.univ-mrs.fr}}
\author{S\'ebastien Descotes-Genon\\
       Laboratoire de Physique Th\'eorique,

CNRS \&Univ. Paris-Sud 11 (UMR 8627), 91405 Orsay Cedex, France\\
      E-mail: \email{Sebastien.Descotes-Genon@th.u-psud.fr}}
\author{Hai-Bo Li\\
        Institute of High Energy Physics, P.O.Box 918, Beijing 100049, China\\
        E-mail: \email{kangxw@ihep.ac.cn}}

\abstract{ In this paper, we exploit the angular and quantum
correlations in the $D\bar{D}$ pairs produced through the decay of
the $\psi(3770)$ resonance in a charm factory to investigate
CP-violation in two different ways. We consider the case of
$\psi(3770)\rightarrow D\bar{D}\rightarrow (V_1V_2)(K\pi)$ decays,
which provide a new way to measure the strong phase difference
$\delta$ between Cabibbo-favored and doubly-Cabibbo suppressed $D$
decays required in the determination of the CKM angle $\gamma$. We
also build CP-violating observables in $\psi(3770)\rightarrow
D\bar{D}\rightarrow (V_1V_2)(V_3 V_4)$ to isolate specific new
physics effects in the charm sector.  Neglecting the systematics, we
give a first rough estimate of the sensitivities of these
measurements at BES-III with an integrated luminosity of 20
fb$^{-1}$ at $\psi(3770)$ peak and at a future super $\tau$-charm
factory with a luminosity of $10^{35}$ cm$^{-2}$s$^{-1}$ }

\FullConference{35th International Conference of High Energy Physics - ICHEP2010,\\
        July 22-28, 2010\\
        Paris France}

\begin{document}

\section{Introduction}

In the framework of standard model (SM), $CP$ violation in the charm
sector is very small, thus any significant amount of $CP$ violation
will be a clean signal of new physics (NP). There have been much
papers on this. In our work, we will fully exploit $D\to VV$ modes
which exhibit rather large branching ratios, of similar size with
respect to $PP$ ($P$ denote pseudoscalar meson) or $VP$ ($V$ denote
vector meson) modes, and provide further new observables. These
points have not been detailed so far and can be verified at BES-III
or other charm factories.

\section{Correlated $D$ decay}

$D^0\overline{D}^0$ pair produced in $\psi(3770)$ is in
antisymmetric coherent state which can be written as
\begin{equation}
|(D\bar{D})_{L=1}\rangle=\frac{-|D_1\rangle|D_2\rangle+|D_2\rangle|D_1\rangle}{\sqrt{2}}.
\end{equation}
For correlated $D$ decays,
one can in principle consider the following different situations,
\begin{itemize}
\item $(PP)+(PP),(PP)+(VP),(VP)+(VP)$: the only available observable is the
branching ratio, since the partial waves and helicities are all
fixed by angular momentum conservation.
\item $(PP)+(VV),(VP)+(VV)$: $(VV)$ can have three helicity states, and thus there are new angular observables.
This can be exploited for $(PP)=K\pi$ in connection with the measurement of the CKM angle $\gamma$.
\item $(VV)+(VV)$: this will be studied with an interest in new observables for CP-violation.
\end{itemize}
Now we list the decay chains, $\psi \to D_1 D_2,\,\,
 D_1 \to V_1 V_2,\,\, D_2 \to K\pi$ for $\gamma$ measurement and $\psi \to D_1
 D_2,\,\,D_1 \to V_1 V_2,\,\, D_2 \to V_3 V_4$ for $CP$ violation, with all the vector mesons
sequential decaying to their pseudoscalars.

Next we will construct the observables from the differential decay
width expressed in helicity angle and helicity amplitudes
corresponding to these two decay chains.

\section{Observables and potential at charm factories}
\subsection{For $\gamma$\, measurement}
Introducing $r\cdot e^{i\delta}=\frac{\langle
K^-\pi^+|\overline{D}_0\rangle}{\langle K^-\pi^+|D_0\rangle}$, the
differential decay width can be written as \cite{DDbar}
\begin{eqnarray}
\label{eq:angdist} d\Gamma_{2V} &=&  \frac{9}{4\pi}
 d(\cos\theta_{V_1}) d(\cos\theta_{V_2}) d\Phi
\times  |A^{\psi V_1 V_2}|^2 |A^{D^0\to K\pi}|^2 \\ \nonumber
&&\times \Big[\cos^2\theta_{V_1}\cos^2\theta_{V_2}
  |A^{D^0\to V_1V_2}_0|^2(1+2r\cos\delta+r^2)\\ \nonumber
&&\quad +\frac{1}{2}\sin^2\theta_{V_1}\sin^2\theta_{V_2} \cos^2\Phi
  |A^{D^0\to V_1V_2}_{||}|^2(1+2r\cos\delta+r^2)\\ \nonumber
&&\quad
-\sqrt{2}\cos\theta_{V_1}\sin\theta_{V_1}\cos\theta_{V_2}\sin\theta_{V_2}
\cos\Phi Re[A^{D^0\to V_1V_2}_0(A^{D^0\to
V_1V_2}_{||})^*](1+2r\cos\delta+r^2)\\&&\quad+\cdots\Big]
\end{eqnarray}
In the above expression, we see that,
\begin{itemize}
 \item The branching ratio only depends on the three amplitude combinations
    \begin{equation}
     M_0 = A_0(1+r e^{i\delta}),\quad
     M_{||} = A_{||}(1+ re^{i\delta}),\quad
     M_\perp = A_\perp(1-re^{i\delta}).
    \end{equation}
 \item since $\delta$ is small, the sensitivity on sine in addition to cosine (PP case) is expected to improve
 the final results.
 \end{itemize}
 Thus, the above constraint can be improved by exploiting the expected knowledge of polarization of VV modes (single-tag),
 then the measurement of $M_i$ in the correlated decay
(double-tag) may lead to a better result on $\delta$.

The error on $\cos\delta$ is given by \cite{Mao-Zhi Yang}
   \begin{equation}
     \Delta (\cos \delta) \approx \frac{1}{2r\sqrt{N_{K^-\pi^+}}}\approx \frac{\pm 284.5}{\sqrt{N(D^0\bar{D}^0)}}.
   \end{equation}
At BES-III, about $72 \times 10^6$ $D^0 \overline{D}^0$ pairs can be
collected with four years running, which implies an accuracy of
about 0.03 for $\cos\delta$, when considering both $K^- \pi^+$ and
$K^+\pi^-$ final states. Citing the present average result of
$\delta=(26.4^{+9.6}_{-9.9})^{\circ}$, we can get the error of
$\delta$, $\Delta(\delta)=\pm3.9^\circ$ at BES-III, and
$\Delta(\delta)=0.4^\circ$ at super-$\tau$-charm factory with the
luminosity about 100 times improvement than BES-III. At this stage,
the results are pure statistics. The true experimental systematics
are required to be studied. Here we want to emphasize one thing
again, size of other terms (e.g. $\sin\delta$ ) has not been studied
yet and expected to improve the measurement.

\subsection{For $CP$\, violation}
 If we take the decay chain \cite{Bigi}
    \begin{equation}
       e^+e^- \to \psi \to D^0 \bar{D}^0 \to f_a f_b
    \end{equation}
 with $f_a$ and $f_b$ CP eigenstates of the same CP-parity, we have
    \begin{equation}
 CP|\psi\rangle = |\psi\rangle \qquad CP|f_af_b\rangle =\eta_a\eta_b(-1)^\ell|f_a f_b\rangle = -|f_a f_b\rangle
    \end{equation}
 since $f_a$ and $f_b$ are in a $P$ wave. Therefore, the decay of $\psi$ into the states of identical $CP$ parity is, by itself, a $CP$ violating
 observable. In fact one can obtain the following combined branching ratio with neglecting $CP$
 violation in $D^0\overline{D}^0$ mixing \cite{Zz},
    \begin{equation}
  Br((D^0\bar{D}^0)_{C=-1} \to f_a f_b)= 2Br(D_0\to f_a)Br(D_0\to f_b)(\left|\rho_a-\rho_b\right|^2+r_D|1-\rho_a\rho_b|^2 )\\
    \end{equation}
 with
    \begin{equation}
     \rho_f=\frac{A(\bar{D}^0\to f)}{A(D^0\to f)},\quad r_D=(x^2+y^2)/2<10^{-4}
    \end{equation}
  Thus,  $CP$ conservation at the level of the amplitude would require that only two combinations of
  transversity amplitudes are allowed: $(0,\perp)$ or $(||,\perp)$ since we know the parallel helicity
  ``||'' is $CP$ even and the perpendicular one
  ``$\perp$'' is $CP$ odd. Other  combinations such as $(0,0)\,(0,||)\,(||,0) \, (||,||)\,
 (\perp,\perp)$ should be $CP$ violating observables. Exploiting orthogonality relationships for Legendre
 and Chebyshev polynomials to select specific angular dependence from the whole differential decay width, one can
 get these $CP$ violating observables \cite{DDbar},
 \begin{eqnarray}
 &&\int  d\Gamma_{4V} \frac{1}{8} (5 \cos^2 \theta_{V_1}-1) (5 \cos^2 \theta_{V_2}-1)(5 \cos^2 \theta_{V_3}-1)(5 \cos^2 \theta_{V_4}-1)
 \nonumber \\ \nonumber
 && \hspace{2cm}=|A^{\psi V_1 V_2 V_3 V_4}|^2 |A^{D_0\to V_1V_2}_0|^2 |A^{D_0\to V_3V_4}_0|^2\times
 |\rho^0_{V_1,V_2}-\rho^0_{V_3,V_4}|^2 \\
 &&\int  d\Gamma_{4V}\frac{1}{32} (5\cos^2 \theta_{V_1}-3) (5 \cos^2 \theta_{V_2}-3) (5 \cos^2 \theta_{V_3}-3) (5 \cos^2
 \theta_{V_4}-3)\nonumber\\
 &&\qquad \quad \cdot(4\cos^2\Phi-1)(4\cos^2\Psi-1)\nonumber\\
 && \hspace{2cm}=|A^{\psi V_1 V_2 V_3 V_4}|^2 |A^{D_0\to V_1V_2}_{||}|^2 |A^{D_0\to V_3V_4}_{||}|^2\times
 |\rho^{||}_{V_1,V_2}-\rho^{||}_{V_3,V_4}|^2\nonumber\\[0.2cm]\nonumber
 &&\cdots
 \end{eqnarray}
 Note that this projection yields CP violating observables without performing a full angular analysis.

 If we parameterize $\rho_f$ as $\rho_f=\eta_f (1+\delta_f) e^{i\alpha_f}$ ($\delta_f$ is CP violation in decay and can be
 negligible.), we can get, as an illustrative example, the branching ratio for the the most promising channel $\rho^0 \rho^0$/$\bar{K}^{*0} \rho^0$ which
 has large branching ratio among the $CP$ eigenstates,
   {\small\begin{eqnarray}
 Br((D^0\bar{D}^0)_{C=-1} \to \rho^0 \rho^0,\bar{K}^{*0} \rho^0)\Big|^{CPV}_{(0,||)} \simeq 8 Br^0(D^0\to\rho^0
 \rho^0)\cdot Br^{||}(D^0\to\bar{K}^{*0} \rho^0) \sin^2\frac{\alpha_a-\alpha_b}{2}.
    \end{eqnarray}}
 ``0'' and ``||'' in the superscript means the corresponding fraction.
 Assuming no $CP$ violating signal events are observed we have the upper
 limit, $|\alpha_a-\alpha_b|<4.4^{\circ}$ at 90\%-C.L. at BESIII and $|\alpha_a-\alpha_b|< 0.5^{\circ}$ at 90\%-C.L. at
 super-$\tau$-charm factory. Its branching fraction
will be estimated to the level of less than $10^{-7}$ if there is no
CP violating events at BES-III. At super $\tau$-charm factory it
would be reduced by one order.

 \section{conclusion}
In the case of CP-tagged $D\rightarrow K\pi$ decays, we expect the
determination of the error on $\delta$ can be improved by taking
into account the dependence of the full angular decay width to the
sine of the strong phase. In the case of $\psi(3770) \rightarrow
D^0\bar{D}^0\rightarrow (V_1V_2)(V_3 V_4)$, CP-violating observables
can be constructed and phase differences are discussed. To conclude,
we say again that a further careful study of experimental
systematics is required since they presumably dominate the quoted
uncertainty here.

\end{document}